\documentclass[prd,preprint,aps,dvips]{revtex4}
\usepackage[english]{babel}
\usepackage{amscd}
\usepackage{epsfig}
\usepackage{tabularx}
\usepackage{graphicx}
\usepackage{latexsym}
\usepackage{amsmath}
\usepackage{amsfonts}
\usepackage{amssymb}
\usepackage[latin1]{inputenc}
\usepackage{times}
\usepackage[T1]{fontenc}
\usepackage{latexsym}
\usepackage{graphics}
\usepackage{verbatim}

\newcommand{\be}{\begin{equation}}
\newcommand{\ee}{ \end{equation}}
\newcommand{\ben}{\begin{eqnarray}}
\newcommand{\een}{\end{eqnarray}}

\begin{document}

\title{$q$-Deformed Landau diamagnetism problem embedded in D-dimensions}

\author{Francisco A. Brito and Andre A. Marinho}

\affiliation{Departamento de F\'\i sica, Universidade Federal de
Campina Grande, 58109-970 Campina Grande, Para\'\i ba, Brazil}

\date{\today}


\begin{abstract} 
We address the issue of 
generalizing the thermodynamic quantities via $q$-deformation, i.e., via the $q$-algebra that describes $q$-bosons and $q$-fermions. In this study with the application of $q$-deformation to the Landau diamagnetism problem in two 
dimensions, embedded in a $D$-dimensional space, we will attempt to get a better understanding of the 
$q$-deformation. We obtain new results for $q$-deformed internal energy, number of particles, magnetization and magnetic susceptibility, which recover the values already known in the literature in the limit $q\to1$.
\end{abstract}


\maketitle


\section{Introduction}

The study of quantum groups and quantum algebras has attracted a great interest in 
recent years and stimulated intense research in several areas of physics \cite {bie}, 
taking into account a wide spectrum of applications, since cosmic strings and black holes 
to the fractional quantum Hall effect and high-Tc superconductors \cite {wil} and theories of 
rational field, non-commutative geometry, the quantum theory of super-algebras and so on \cite{chai}. 
There is no satisfactory universally recognized definition of a quantum group. 

The concept of quantum groups was motivated by problems from a large number of physical situations. The so-called $q$-deformed algebra \cite{gall} has been the object of interest in
literature both in physics and mathematics over the past years. A great effort has been devoted to 
their understanding and development \cite{wil}. One of its main ingredients is a measure of deformation $q$, 
introduced in the commutation relations that define the Lie algebra of the system with the condition that the 
original Lie algebra, not deformed, is produced at the limit $q\to 1$.
From the seminal work of Biedenharn \cite{bie1} and Macfarlane \cite{mac}, it is clear that the q-calculation 
initially introduced at the beginning of the last century by Jackson \cite{jac} based on the study of the hypergeometric 
function \cite{ext}, plays a central role in the quantum group representations with a deep physical significance. 
Indeed, it was shown that 
the $q$-deformed oscillators using the Jackson derivative (JD) or the so-called $q$-derivative operators 
\cite{flo} define a generalized $q$-deformed dynamic of the $q$-commutative phase space. Thus, it appeared recently a great interest in investigating the $q$-deformed thermodynamical systems \cite{lav}.

One possible mechanism capable of generating a deformed version of classical statistical mechanics is to replace the Gibbs-Boltzmann 
distribution,  
by postulating a deformed entropy involving a generalized 
thermodynamic theory. Thus, some generalizations of statistical mechanics were proposed \cite{tsa,abe, kan, kan1}. It has also been demonstrated in \cite{lav1} that a natural realization of thermodynamics of $q$-deformed bosons and fermions can be found in the formalism of $q$-calculus. In fact, it was shown that the q-integration is related to the free energy of spin systems \cite{erz} --- see also \cite{bim}.

In a more specific case, we focus our attention to studying of the Landau diamagnetism problem. 
The Landau diamagnetism problem continues to raise issues that have strong relevance today. These issues are related to the 
inherent quantum nature of the problem. 
It can be used as a phenomenon to illustrate the essential role of quantum mechanics in the 
surface and perimeter corrections, the dissipation of statistical mechanics of non-equilibrium, and others. 

The paper is organized as follows. In Sec.~\ref{II} we present the q-deformed algebra and Jackson derivative. In Sec.~\ref{III} we develop
the q-deformed Landau diamagnetism. In Sec.~\ref{IV} we make our final comments.

\section{$q$-Deformed quantum algebra}
\label{II}

The $q$-deformed algebraic symmetry of the quantum oscillator is defined by the $q$-deformed Heisenberg algebra in terms of 
creation and annihilation operators $c^{\dagger}, c$ and $N$ as \cite {chai, ng, sak}
\begin{eqnarray} [c,c]_k = [c^{\dagger},c^{\dagger}]_k =0\quad,\qquad\qquad\ cc^{\dagger} -kqc^{\dagger}c=q^{-N},\end{eqnarray}
\begin{equation}[N,c^{\dagger}]= c^{\dagger}\quad, \qquad\qquad [N,c] = -c ,\end {equation}
where the deformation parameter $q$ is real, being the constant $k = 1$ for $q$-bosons (with commutators) and $k = -1 $
for $q$-fermions (with anticommutators). 
The basic $q$-deformed quantum number is defined as
\begin{equation}\label{eq8} [x]=\frac {q^{x}-1} {q-1}.\end{equation}
In addition, the operators obey the relations
\begin{eqnarray}[x,y]_k=xy-kyx, \qquad\qquad cc^{\dagger}=[1+kN].\end {eqnarray}
Note that for $q\neq1$ the $q$-deformed quantum number $[x]$ does not meet additivity
\begin{equation} [x+y] = [x] + [y]+(q-1)[x][y],\end{equation}
whereas in the limit of $q\to 1$ the basic $q$-deformed quantum number $[x]$ is reduced to an ordinary number
$x$. 

The $q$-Fock space spanned by the orthornormalized eigenstates $|n>$ is constructed according to
\begin{eqnarray} {|n>} =\frac{(c^{\dagger})^{n}} {\sqrt{[n]!}}{|0>}\quad,\qquad\qquad a{|0>}=0 ,\end {eqnarray}
where the factorial of the basic $q$-deformed quantum number $[n]$ is defined as
\begin{equation}[n]!=[n][n-1].....[1].\end{equation}
The actions of $c$, $c^{\dagger}$ and $N$ on the states $|n>$ in the $q$-Fock space are known to be
\begin{equation} c^{\dagger}{|n>} = [n+1]^{1/2} {|n+1>},\end{equation}
\begin{equation} c{|n>} = [n]^{1/2} {|n-1>},\end{equation}
\begin{equation} N{|n>} = n{|n>}.\end{equation}
One may transform the $q$-Fock space into the configuration space (Bargmann holomorphic representation) \cite{flo,fin} as in the following
\begin{eqnarray} c^{\dagger} = x,\qquad\qquad c = D_x^{(q)},\end{eqnarray}
where $D_x^{(q)}$ is the Jackson derivative (JD) \cite{jac}
\begin{equation}D_x^{(q)}f{(x)}=\frac {f{(qx)}- f{(x)}}{x{(q-1)}}.\end{equation}
Note that it becomes an ordinary derivative as $q=1$. Therefore, JD naturally occurs in quantum deformed structures.
It turns out to be a crucial role in $q$-generalization of thermodynamical relations \cite{lav1}.

\section {$q$-Deformed Landau diamagnetism problem embedded in $D$-dimensions}
\label{III}

To explain the phenomenon of diamagnetism, we have to take into account the interaction between the external 
magnetic field and the orbital motion of electrons. Disregarding the spin, the Hamiltonian of a particle of mass 
$m$ and charge $e$ in the presence of a magnetic field \textbf{H}, is given by the expression 
\begin{equation} H=\frac {1}{2m} \left(\textbf{p}-\frac{e}{c}\textbf{A}\right)^2,\end{equation}
where \textbf{A} is the vector potential associated with the magnetic field $\textbf{H}$ and c is the speed of light. 
Let us start to formalize the statistical mechanical problem by using the grand partition function in the form
\begin{eqnarray}\ln{\Xi}&=&-k\frac{2eHL^{2}}{{hc}}\displaystyle\sum_{n=0}^{\infty}\frac{L^{D-2}}{(2\pi)^{D-2}}\int_{-\infty}^{\infty}
{d^{D-2}}k_z\ln\Big\{{1-kz\:{\exp}}\Big[-\frac{\beta\hbar^{2}k_z^{2}}{2m}\nonumber\\&-&\frac{\beta\hbar
eH}{mc}\Big(n+\frac{1}{2}\Big)\Big]\Big\},\end{eqnarray}
where
$k_z^{2}=k_1^{2}+k_2^{2}+....+k_{(D-2)}^{2}$. Recall that $k=\pm1$ for bosons and fermions, respectively. Here, we have generalized the Landau diamagnetism problem,
usually embedded in three dimensions, to embed it in $D-$dimensions, i.e., we have now $D-2$ transversal dimensions. 
Below, we further generalize this result by also applying the Jackson derivative. 

However, our study is focused on the analysis of diamagnetism in the limit of high temperatures
$(z\ll1)$ where we apply the $q$-deformed algebra. Thus, the partition function is written as follows
\begin{equation}\label{eq}{\ln}\Xi=\frac{2eHL^{D}}{\lambda^{(D-2)}{(2\pi)}^{(D-2)}hc}z\displaystyle\sum_{n=0}^{\infty}y_n ,\end{equation}
where $z=\exp(\beta\mu)$ is the fugacity, $y_n=\exp\left[\left(\frac{-\beta\hbar
eH}{mc}\right)\left(n+\frac{1}{2} \right)\right]$, $\lambda=\frac{\hbar}{(2\pi m\kappa_BT)^{1/2}}$ 
is the thermal wavelength. For the sake of simplicity, in our calculations we simply use 
$y_n=\exp(-\beta\epsilon_n)$, where $\epsilon_n=\frac{\hbar
eH}{mc}\left(n+\frac{1}{2}\right)$. With these substitutions we 
can rewrite the equation (\ref{eq}) as
\begin{equation} {\ln{\Xi}}=zC\displaystyle\sum_{n=0}^{\infty}y_n, \qquad C=\frac{2L^{(D)}eH}{\lambda^{(D-2)}(2\pi)^{(D-2)} hc}.\end{equation}

Now we are ready to applying the $q$-deformed algebra
via the introduction of JD, which is a key factor of $q$-deformed thermodynamics.
This is achieved through the modification of the ordinary thermodynamical derivatives as follows \cite{lav1}
\begin{equation}\label{JD0} 
\frac{\partial}{\partial{z}}\to D_z^{(q)},\qquad \frac{\partial}{\partial{y_i}}\to D_{y_i}^{(q)},\end {equation}
where $D_z^{(q)}$ and $D_{y_i}^{(q)}$ are the JD's.
\subsection{\textbf{$q$-Deformed thermodynamical quantities}}

The $q$-deformed algebra earlier discussed leads to $q$-deformed statistical mechanics of the quantum oscillator whose deformation is encoded in the occupation number
$n^{(q)}_i$, say for bosons \cite{lav1} 
\begin{eqnarray}
n^{(q)}_i=\frac{1}{\ln{q}}\ln{\left(\frac{z^{-1}\exp{(\beta\epsilon_i)}-1}{z^{-1}\exp{(\beta\epsilon_i)}-q}\right)},
\end{eqnarray}
with the $q$-deformed number now defined as 
\begin{eqnarray}\label{eqI}
N^{(q)}=\sum_i n^{(q)}_i\equiv zD_z^{(q)}\ln \Xi,
\end{eqnarray}

Because we are working with $z\ll1$ in our previous partition function we also apply this limit to $n^{(q)}_i$ to find the leading term
\begin{eqnarray} n^{(q)}_i=\frac{q-1}{\ln{q}}n_i, \qquad n_i=z\exp(-\beta\epsilon_i), \end{eqnarray}
such that 
\begin{eqnarray}\label{eqII}
N^{(q)}=\frac{q-1}{\ln{q}}N, \qquad N=\sum_i n_i\equiv z\frac{\partial}{\partial z}\ln \Xi.
\end{eqnarray}
The same reasoning applies to the internal energy so that we can use Eqs.~(\ref{eqI}) and (\ref{eqII}) to establish the following relations
\begin{eqnarray}\label{JD}
D_z^{(q)}=\frac{q-1}{\ln{q}}\frac{\partial}{\partial z}, \qquad D_{y_i}^{(q)}=\frac{q-1}{\ln{q}}\frac{\partial}{\partial y_i}.
\end{eqnarray}

We shall now apply these results to our specific example. As one knows the internal energy is defined in the form
\begin{equation} U= -\frac {\partial}{\partial{\beta}}\ln{\Xi}.\end{equation}
Thus, to implement the JD is necessary first to change the basis as follows
\begin{equation}-\frac {\partial}{\partial{\beta}} = -\left(\frac{\partial{y_n}}{\partial\beta}
\frac{\partial}{\partial{y_n}}\right),\end{equation}
such that
\begin{equation} U=-\frac{\partial{y_n}}{\partial\beta}D_{y_n}^{(q)} \ln{\Xi} ,\end{equation}
\begin{equation} U= -zC\displaystyle\sum_{n=0}^{\infty}\frac{\partial{y_n}}{\partial\beta}D_{y_n}^{(q)}\, y_n.\end{equation}
So using the JD definition in Eq.~(\ref{JD0}) and its approximation given in Eq.~(\ref{JD}) we find
\begin{equation} U=-zC\frac{q-1}{\ln {q}}\frac{\partial}{\partial\beta}\displaystyle\sum_{n=0}^{\infty}y_n.\end{equation}
Now performing the summation we find the following
\begin{equation}\displaystyle\sum_{n=0}^{\infty}y_n=\frac{1}{2\sinh(\beta\mu_BH)},\end{equation}
where $\mu_B=\frac{\hbar e}{2mc}$. Therefore the internal energy is written as
\begin{equation}U=zC\frac{q-1}{\ln {q}}\left(\frac{\mu_B\cosh(\beta\mu H)}{2\sinh^{2}(\beta\mu H)}\right).\end{equation}
To calculate the number of particles we have
\begin{equation}\label {eq2}N=z\frac{\partial}{\partial z}\ln{\Xi}.\end{equation}
Now applying into Eq.~(\ref {eq2}) the JD, using the JD definition in Eq.~(\ref{JD0}) and its approximation given in Eq.~(\ref{JD}), just as we did for the internal energy case, we get
\begin{equation}\label{eq2.1}N=zC\frac{q-1}{\ln {q}}\frac {1}{2\sinh{(\beta\mu_B H)}}.\end{equation}

\subsection{\textbf {The specific heat}}

To determine the specific heat, we first make use of the Jacobian
\begin{equation}\label{eq3}C_V=\frac {1}{N}\left(\frac{\partial U}{\partial T}\right)_{V,N} =
\frac{-\kappa_B\beta^{2}}{N}\left(\frac{\partial U}{\partial\beta}\right)_{V,N},\end{equation}
\begin{eqnarray}\Big(\frac{\partial U}{\partial\beta}\Big)_N &=& \frac{\partial(U,N)}{\partial(\beta,N)}=
\frac{\partial(U,N)}{\partial(\beta,z)}\frac{\partial(\beta,z)}{\partial(\beta,N)}\nonumber\\&=&
\Big(\frac{\partial U}{\partial\beta}\Big)_z -\Big(\frac{\partial U}{\partial z}\Big)_\beta\frac{\Big(\frac{\partial
N}{\partial\beta}\Big)_z}{\Big(\frac{\partial N}{\partial z}\Big)_\beta}.\end{eqnarray}
Using the derivatives above, we find
\begin{equation}\label{eq4}\left(\frac{\partial U}{\partial\beta}\right)_N = \frac{-zC}{2}\frac{q-1}{\ln{q}}\frac{\mu_B^{2}H}
{\sinh^{3}(\beta\mu_B H)}.\end{equation}
Applying the equation (\ref{eq4}) in the equation (\ref{eq3}) and substituting $N$
obtained in the equation (\ref{eq2.1}), we arrive at the specific heat
\begin{equation}C_V=\kappa_B\left(\frac{\beta\mu_B H}{\sinh(\beta\mu_B H)}\right)^{2}.\end{equation}
Recall that $\kappa_B$ is the Boltzmann constant. This coincides with the usual result.

\subsection{\textbf {$q$-Deformed magnetization and susceptibility}}

For determining the magnetization we carried out the thermodynamical derivative
\begin{equation}M=-\frac{\partial\phi}{\partial H},\end{equation}
where the grand potential $\phi$ is determined as
\begin{eqnarray}\phi=-\frac{1}{\beta}\ln{\Xi}\quad, \qquad\qquad \phi=\frac {-zC}{2\beta\sinh(\beta\mu_B H)}.\end{eqnarray}
However, to implement the JD we have to make the change as follows
\begin{equation}-\frac{\partial \phi}{\partial H}=-\left(\frac{\partial y_n}{\partial H}\frac{\partial \phi} {\partial
y_n}\right).\end{equation} 
Thus we find 
\begin{equation} M=- \frac{\partial y_n}{\partial H}D_{y_n}^{(q)} \phi ,\end{equation} and
\begin{equation} M=\frac {z{C^{*}}}{\beta} \frac{q-1}{\ln {q}} \frac {\partial}{\partial H}\left(2H\displaystyle\sum_{n=0}^{\infty}y_n
\right),\end{equation}
where $C^*=C/2H$, such that
\begin{eqnarray} M=\frac {z{C^{*}}}{\beta} \frac {q-1}{\ln {q}} \frac{1}{\sinh(\beta\mu_B H)}
\left(1- \frac{\beta\mu_B H\cosh(\beta\mu_B H)}{\sinh(\beta\mu_B H)}\right).\end{eqnarray}
We can also eliminate the chemical potential through the number of particles $N$ and rewrite the
magnetization as 
\begin{equation} M = \frac{N}{\beta H} \left(1- \frac{\beta\mu_B H\cosh(\beta\mu_B H)}{\sinh(\beta\mu_B H)}\right),\end{equation}
or in terms of the Langevin  function given by
\begin{equation}{\cal L}(\beta\mu_B H)= \coth(\beta\mu_B H)-\frac{1}{\beta\mu_B H},\end{equation}
we get to
\begin{equation}\label{eq5} M=-N \mu_B{\cal L}(\beta\mu_B H), \end{equation} which is formally the same as the usual magnetization, with
$N$ playing the role of $q$-deformed number of particles.

Let us now make the analysis in the weak field limit $\beta\mu_B H\ll 1$ as follows.
The use of this limit into the equation (\ref{eq5}) give us
\begin{equation}M=-\frac{N\mu_B^{2}H}{3\kappa_B T}.\end{equation}
Now computing the susceptibility reads
\begin{eqnarray}\label{eq-chi} \chi =\frac{\partial M}{\partial H}= -\frac{N_0\mu_B^{2}}{3\kappa_B T},\end{eqnarray} 
where $N_0=z C^* \frac{q-1}{\ln{q}}$. Since $C^*\propto \left(\frac{L}{2\pi\lambda}\right)^D4\pi^2\lambda^2$ one can find a relation between 
$q$ and the number of extra dimensions $n$ as follows. Consider $D=3+n$, and $L^D=L^{3+n}$. If the size of the extra dimensions is not necessarily the same as the size of the three spatial dimensions of the visible Universe, e.g., $\ell\ll L$ such that $L^D\to L^3\ell^n$, then we can write $N_0$ as 
\begin{eqnarray}\label{eq-N0}
N_0=z\frac{q-1}{\ln{q}}C^*\propto zG(q,\ell,n,\lambda)\frac{L^{3}}{2\pi\lambda},\qquad G(q,\ell,n,\lambda)=\frac{q-1}{\ln{q}}\left(\frac{\ell}{2\pi\lambda}\right)^{n},
\end{eqnarray}
where $G(q,\ell,n,\lambda)$ measures the new `strength' of the susceptibility. For a fixed value found $G^*$ one allows us to write the relations
\begin{eqnarray}
n = \ln\left(\frac{G^*\ln{q}}{q-1}-\frac{\ell}{2\pi\lambda}\right),\qquad \frac{\ell}{2\pi\lambda} \leq \frac{G^*\ln{q}}{q-1}-1.
\end{eqnarray}
It is expected that for $L\gg\ell$ one may find $G(q,\ell,n,\lambda)=1$ (when $q\to1$) and for $L\sim\ell$ one has $G(q,L,n,\lambda)\neq1$ which may change with the size $L$ of a sample of a material in a way that goes like $L^{3+n}$. Notice that $0<q<1$ can play the role of a scale that brings the size of extra dimensions higher even if $G^*$ approaches unit. 

In summary, from Eqs.~(\ref{eq-chi}) and (\ref{eq-N0}) we see that for $D=3$ (the usual case) 
the susceptibility varies with the material size according to the power law $L^3$, such that deviations to a power law $L^{3+n}$ may reveal a signature of extra dimensions. One may find in the literature experimental results for susceptibility changing with the
size of a material sample, see for instance \cite{thompson}.

\begin{figure}[ht]
\centerline{
\includegraphics[{angle=90,height=5.0cm,angle=270,width=5.0cm}]{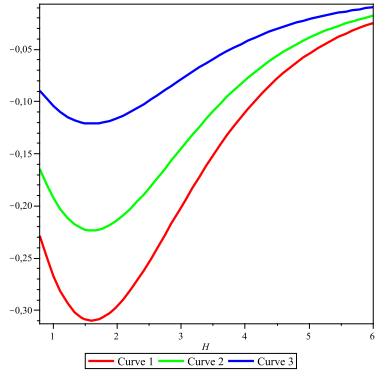}
}\caption{$q$-deformed magnetization as a function of magnetic field $H$ for $q=0.1$ (blue), $q=0.5$ (green), $q=0.99$ (red).}\label{figur2}
\end{figure}

\begin{figure}[ht]
\centerline{
\includegraphics[{angle=90,height=5.0cm,angle=270,width=5.0cm}]{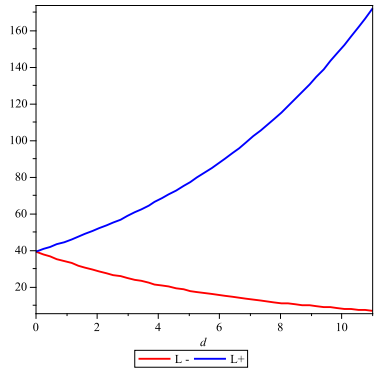}
}\caption{$C^*$ as a function of the number of dimensions for $L_-=2\pi-0.9$ (red), $L_+=2\pi+0.9$ (blue),  $\lambda=1$.}\label{figur3}

\end{figure}

In Fig.~\ref{figur2} is depicted the behavior of the $q$-deformed magnetization for $D=3$ and some values for $q$. The Fig.~\ref{figur3} shows the behavior of $C^*$ as a function of the number of $D$-dimensions for $q\to1$.

\section{\textbf{Conclusions}}
\label{IV}
We apply the $q$-deformed quantum algebra in the Landau diamagnetism problem embedded in $D$-dimensions, in the limit of high temperatures.
We obtain $q$-deformed values $q\neq 1$ for internal energy, number of particles, magnetization, 
susceptibility and specific heat. We see that the q-deformed thermodynamical quantities change, except the specific ones, i.e., those quantities per particles, such as specific heat. That means that all effect of the q-deformation is completely stored into the number of particles $N$. It seems to
be related to {\it impurities} in a sample of a diamagnetic material. As expected, in the limit $q=1$ we see that the results are identical to those in the literature \cite{patt,hua}. The thermodynamical quantities also depend on the number of $D$-dimensions since it is present in the constant $C^*\propto \left(\frac{L}{2\pi\lambda}\right)^D4\pi^2\lambda^2$ --- see Fig.~\ref{figur3}. Note that for $L<2\pi\lambda$ this constant decreases as the number of dimensions $D$ increases. On the other hand, $C^*$ increases with $D$ as $L>2\pi\lambda$, and has no changes as $L=2\pi\lambda$. An interesting phenomenon, for instance, can be achieved in the second case (the bulk sample is larger than the thermal wavelength $\lambda$), where the the magnetization is minimized for $D=3$ (we are assuming all times $D\geq3$). On the other hand, in the first case (the bulk sample is smaller than the thermal wavelength $\lambda$) the magnetization is minimized just for a large number of dimensions. This suggests a mechanism that can be used to select the maximal number of dimensions of the spacetime with minimal diamagnetism, as in our Universe. This fact may find some applications in extra-dimensional physics such as modern cosmology, particle physics and string theory.

\acknowledgments

We would like to thank CNPq, CAPES, and PNPD/PROCAD -
CAPES, for partial financial support.


\end{document}